\documentclass{article}
\usepackage{graphicx} 
\usepackage{tikz}

\usepackage{amsmath}
\usepackage{amssymb}
\usepackage{hyperref}
\usepackage{fullpage}

\usepackage[dvipsnames]{xcolor}

\newcommand{\tikzcircle}[2][red,fill=red]{\tikz[baseline=-0.5ex]\draw[#1,radius=#2] (0,0) circle ;}%

\title{The GHZ Game in Black and White}

\author{Terry Rudolph \\ Imperial College, London, UK \and Tony Newton \\ Newton Astro Labs, London, UK \and Dan-Adrian German \\ Indiana University Bloomington, IN, USA}

\date{July 2026}

\begin{document}

\maketitle

\begin{abstract}
We show that it is possible for three people to win the GHZ game with just the standard real valued misty state formalism, but each party 
will need to have two balls instead of one. That is, there will be six balls involved, rather than three. In fact, as a general 
rule, instead of using complex numbers in a quantum computer you can always add a qubit to ``tag'' complex numbers. But for proofs 
of non-locality, like the GHZ game, we cannot have a single universal tagging qubit, because all three parties would need access to
it and therefore they could potentially transfer information via it. That's where we need to replace the one ball with three that
each only need to be accessed locally.
\end{abstract}

\section{Preliminaries}

This paper was initiated by the first author in 2023 as an alternative to the analysis presented in the last section of \cite{semantics}. 
We assume  the reader is familiar with \cite{terry:01} and \cite{semantics}. Here's our playground for this GHZ game:
\begin{figure}[h]
    \centering
    \includegraphics[scale=0.17]{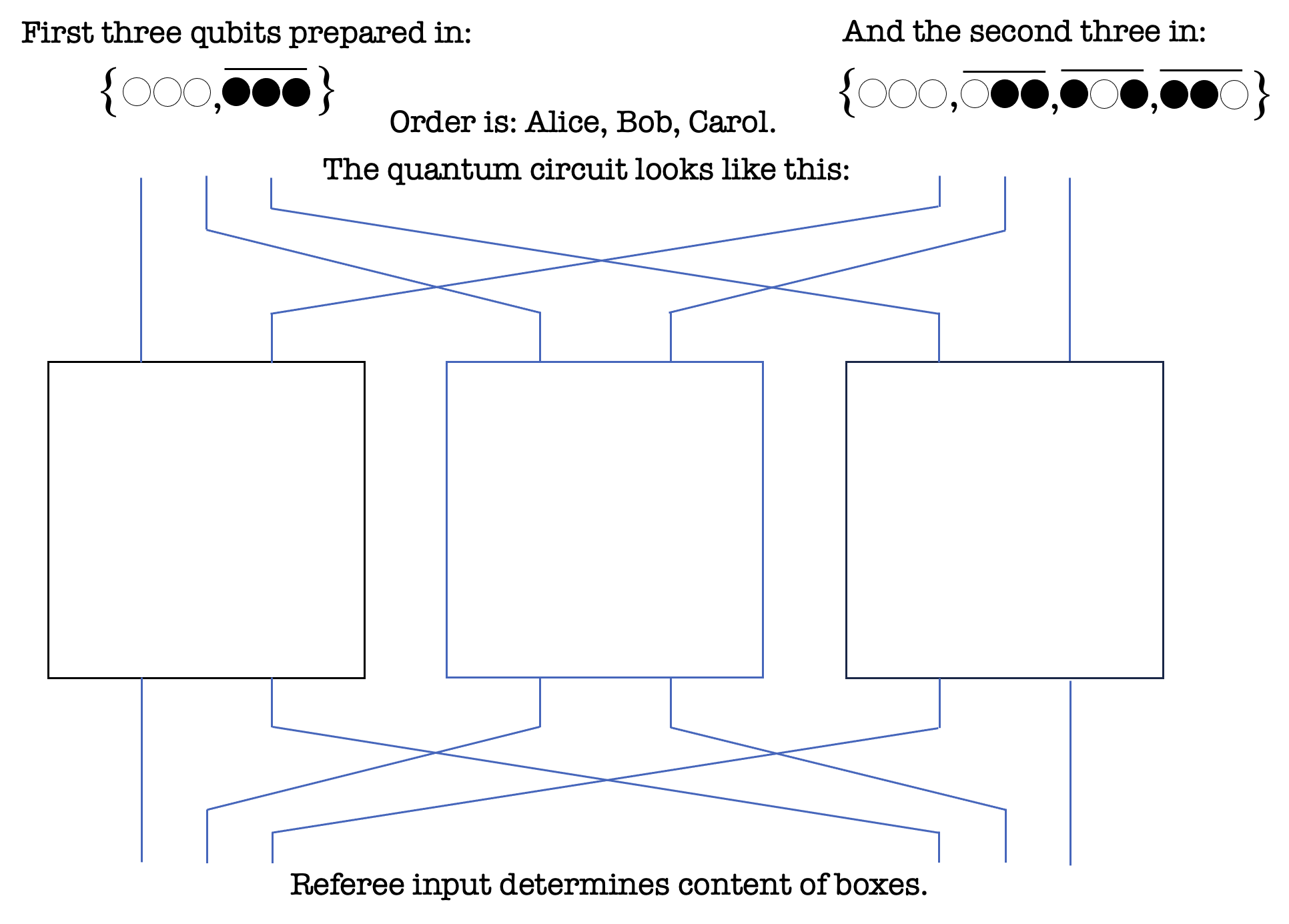}
    \label{fig:ghz-bw-01}
\end{figure}

As a reminder $-\tikzcircle[black, fill=white]{5pt}\tikzcircle[black, fill=black]{5pt}\tikzcircle[black, fill=white]{5pt} = \overline{\tikzcircle[black, fill=white]{5pt}\tikzcircle[black, fill=black]{5pt}\tikzcircle[black, fill=white]{5pt}} = \overline{\tikzcircle[black, fill=white]{5pt}}\tikzcircle[black, fill=black]{5pt}\tikzcircle[black, fill=white]{5pt}=\tikzcircle[black, fill=white]{5pt}\overline{\tikzcircle[black, fill=black]{5pt}}\tikzcircle[black, fill=white]{5pt} = \tikzcircle[black, fill=white]{5pt}\tikzcircle[black, fill=black]{5pt}\overline{\tikzcircle[black, fill=white]{5pt}}$ for $n = 3$ random qubits. Meanwhile, 
the superposition operator describes a set of 
outcomes, so: $- \{\tikzcircle[black, fill=white]{5pt}, \tikzcircle[black, fill=black]{5pt}\} = \overline{\{\tikzcircle[black, fill=white]{5pt}, \tikzcircle[black, fill=black]{5pt}\}} = \{\overline{\tikzcircle[black, fill=white]{5pt}}, \overline{\tikzcircle[black, fill=black]{5pt}}\}$. 

\section{The Game}

Depending on the referee the players will receive one of these four inputs: $000$, $011$, $101$, or $110$. Players who receive a $0$ need to make a measurement
in the $X$ basis (using what we call an $X${\footnotesize -BOX}). 

\begin{figure}[h]
    \centering
    \includegraphics[scale=0.16]{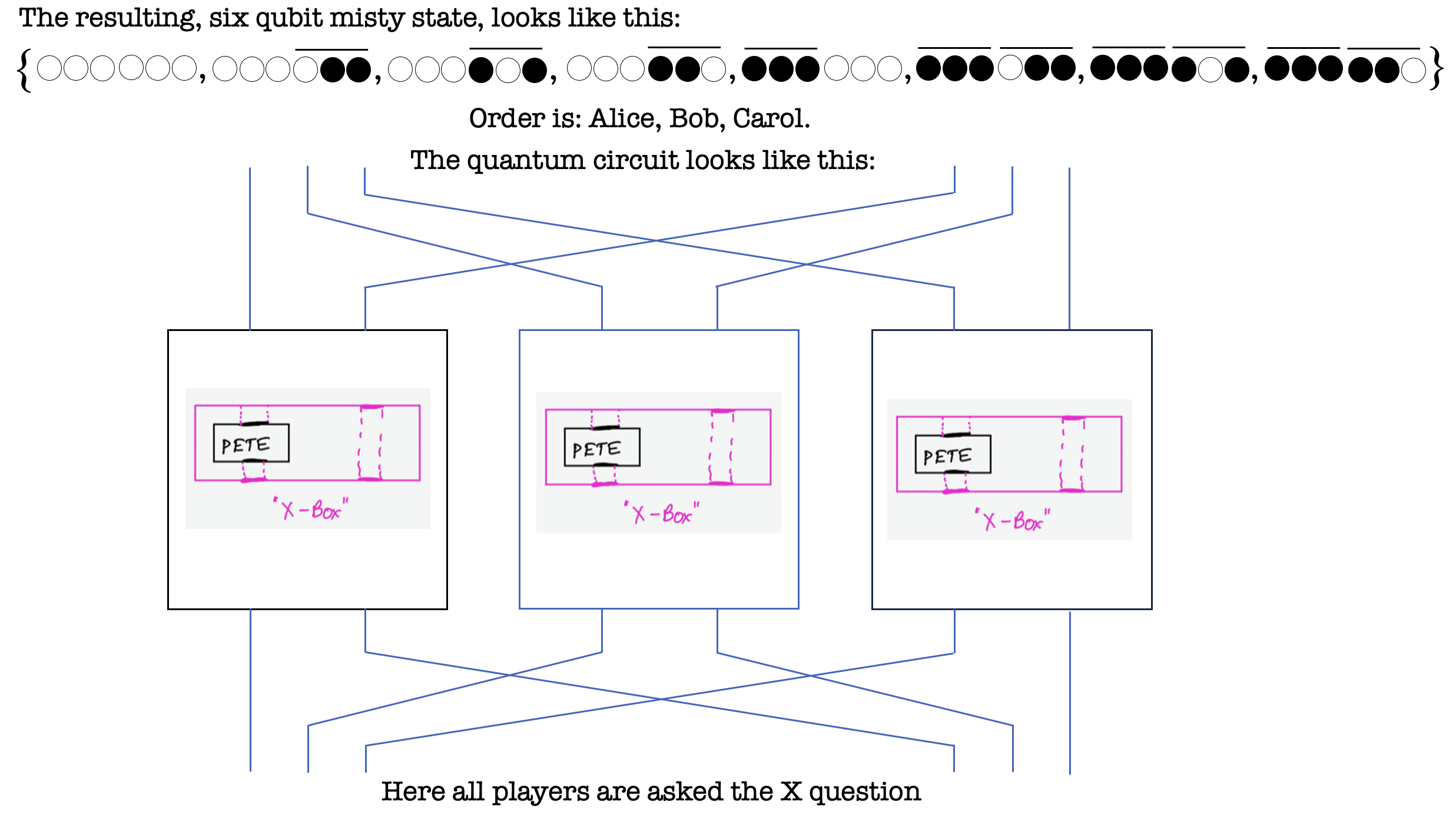}
    \label{fig:ghz-bw-02}
\end{figure}

Players who receive a $1$ will need to make a measurement in the $Y$ basis (using a $Y${\footnotesize -BOX}).

\begin{figure}[h]
    \centering
    \includegraphics[scale=0.16]{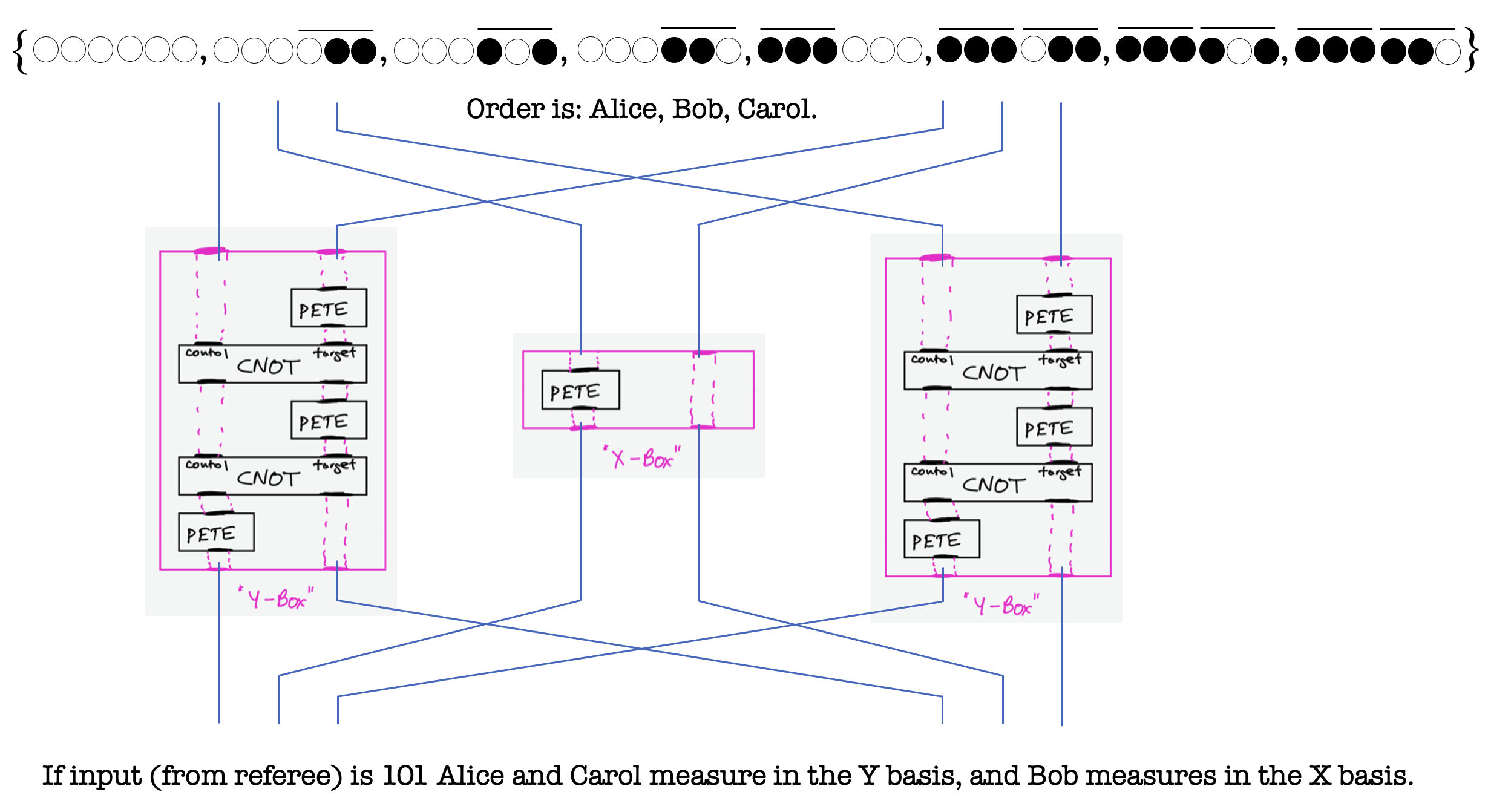}
    \label{fig:ghz-bw-03}
\end{figure}

We will first calculate for the case where all players receive the same question. Then we will consider the
second setup (with Alice and Carol being asked $Y$ and Bob being asked $X$). By symmetry we won't need to consider
any (of the) other cases. We have $8 + 8 = 16$ cases to calculate in total. If we're careful it's not going to be too hard. We start
with the case where all players receive the same input (question).

\newpage 

\section{Calculations (Part I)}

This is how the calculation  of the first input proceeds:

\begin{figure}[h]
    \centering
    \includegraphics[scale=0.19]{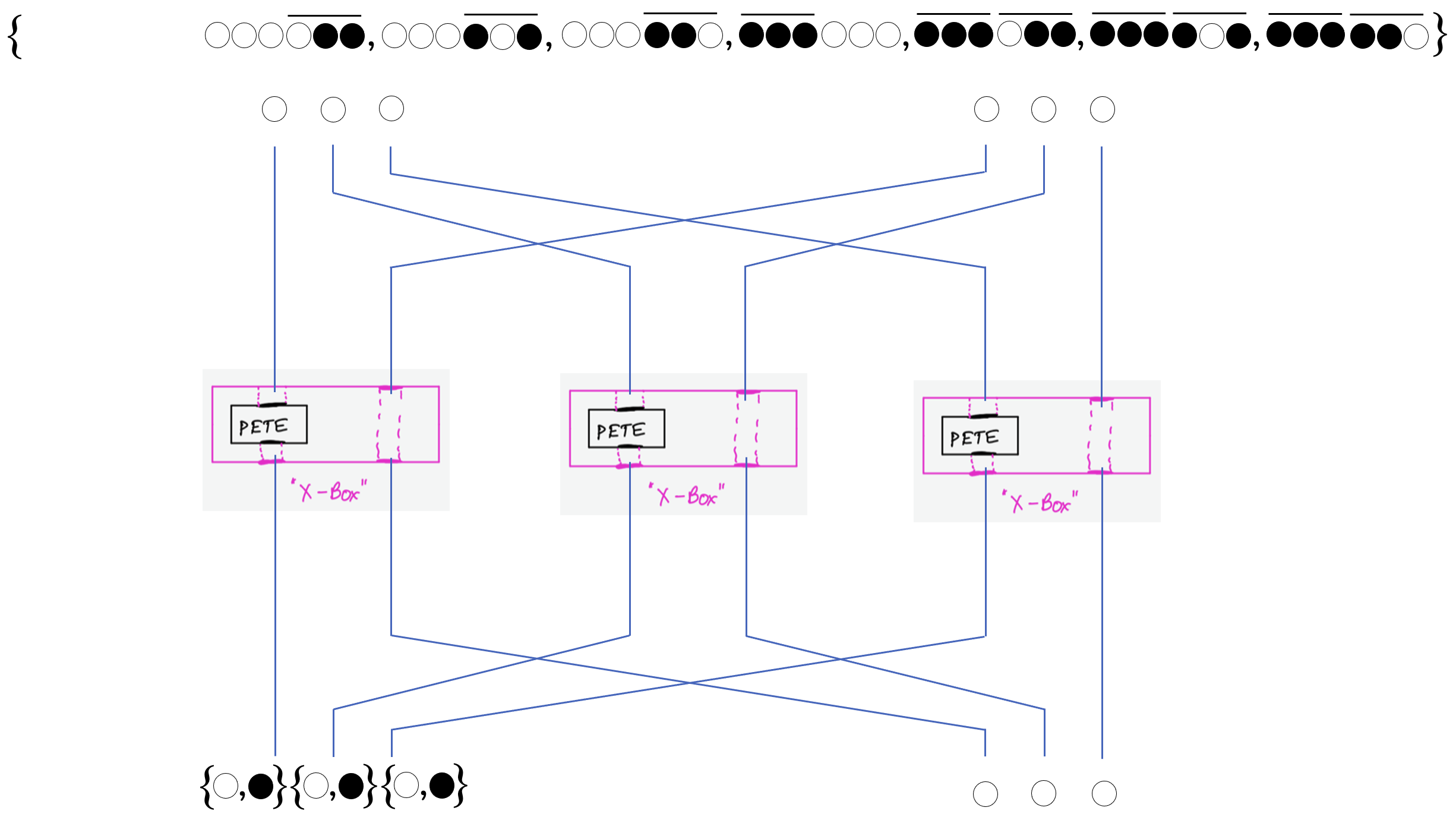}
    \label{fig:ghz-bw-04}
\end{figure}

So now we can state: 

$$\texttt{GHZ}(\tikzcircle[black, fill=white]{5pt}\tikzcircle[black, fill=white]{5pt}\tikzcircle[black, fill=white]{5pt}\tikzcircle[black, fill=white]{5pt}\tikzcircle[black, fill=white]{5pt}\tikzcircle[black, fill=white]{5pt}) = \{\tikzcircle[black, fill=white]{5pt},\tikzcircle[black, fill=black]{5pt}\}\{\tikzcircle[black, fill=white]{5pt},\tikzcircle[black, fill=black]{5pt}\}\{\tikzcircle[black, fill=white]{5pt},\tikzcircle[black, fill=black]{5pt}\}\tikzcircle[black, fill=white]{5pt}\tikzcircle[black, fill=white]{5pt}\tikzcircle[black, fill=white]{5pt} \texttt{}{~~~~~~~~~~~~~~~~~~~~~~~~(1)}$$

Likewise:

$$\texttt{GHZ}(\tikzcircle[black, fill=white]{5pt}\tikzcircle[black, fill=white]{5pt}\tikzcircle[black, fill=white]{5pt}\tikzcircle[black, fill=white]{5pt}\overline{\tikzcircle[black, fill=black]{5pt}}\tikzcircle[black, fill=black]{5pt}) = \{\tikzcircle[black, fill=white]{5pt},\tikzcircle[black, fill=black]{5pt}\}\{\tikzcircle[black, fill=white]{5pt},\tikzcircle[black, fill=black]{5pt}\}\{\tikzcircle[black, fill=white]{5pt},\tikzcircle[black, fill=black]{5pt}\}\tikzcircle[black, fill=white]{5pt}\overline{\tikzcircle[black, fill=black]{5pt}}\tikzcircle[black, fill=black]{5pt} \texttt{}{~~~~~~~~~~~~~~~~~~~~~~~~(2)}$$
$$\texttt{GHZ}(\tikzcircle[black, fill=white]{5pt}\tikzcircle[black, fill=white]{5pt}\tikzcircle[black, fill=white]{5pt}\tikzcircle[black, fill=black]{5pt}\overline{\tikzcircle[black, fill=white]{5pt}}\tikzcircle[black, fill=black]{5pt}) = \{\tikzcircle[black, fill=white]{5pt},\tikzcircle[black, fill=black]{5pt}\}\{\tikzcircle[black, fill=white]{5pt},\tikzcircle[black, fill=black]{5pt}\}\{\tikzcircle[black, fill=white]{5pt},\tikzcircle[black, fill=black]{5pt}\}\tikzcircle[black, fill=black]{5pt}\overline{\tikzcircle[black, fill=white]{5pt}}\tikzcircle[black, fill=black]{5pt} \texttt{}{~~~~~~~~~~~~~~~~~~~~~~~~(3)}$$
$$\texttt{GHZ}(\tikzcircle[black, fill=white]{5pt}\tikzcircle[black, fill=white]{5pt}\tikzcircle[black, fill=white]{5pt}\tikzcircle[black, fill=black]{5pt}\overline{\tikzcircle[black, fill=black]{5pt}}\tikzcircle[black, fill=white]{5pt}) = \{\tikzcircle[black, fill=white]{5pt},\tikzcircle[black, fill=black]{5pt}\}\{\tikzcircle[black, fill=white]{5pt},\tikzcircle[black, fill=black]{5pt}\}\{\tikzcircle[black, fill=white]{5pt},\tikzcircle[black, fill=black]{5pt}\}\tikzcircle[black, fill=black]{5pt}\overline{\tikzcircle[black, fill=black]{5pt}}\tikzcircle[black, fill=white]{5pt} \texttt{}{~~~~~~~~~~~~~~~~~~~~~~~~(4)}$$
$$\texttt{GHZ}(\tikzcircle[black, fill=black]{5pt}\tikzcircle[black, fill=black]{5pt}\tikzcircle[black, fill=black]{5pt}\tikzcircle[black, fill=white]{5pt}\overline{\tikzcircle[black, fill=white]{5pt}}\tikzcircle[black, fill=white]{5pt}) = \{\tikzcircle[black, fill=white]{5pt},\overline{\tikzcircle[black, fill=black]{5pt}}\}\{\tikzcircle[black, fill=white]{5pt},\overline{\tikzcircle[black, fill=black]{5pt}}\}\{\tikzcircle[black, fill=white]{5pt},\overline{\tikzcircle[black, fill=black]{5pt}}\}\tikzcircle[black, fill=white]{5pt}\overline{\tikzcircle[black, fill=white]{5pt}}\tikzcircle[black, fill=white]{5pt} \texttt{}{~~~~~~~~~~~~~~~~~~~~~~~~(5)}$$
For these inputs we took the phase and placed it on the fifth qubit (Bob's second).
Now for the last three cases the phases on our inputs cancel and we have:
$$\texttt{GHZ}(\tikzcircle[black, fill=black]{5pt}\tikzcircle[black, fill=black]{5pt}\tikzcircle[black, fill=black]{5pt}\tikzcircle[black, fill=white]{5pt}\tikzcircle[black, fill=black]{5pt}\tikzcircle[black, fill=black]{5pt}) = \{\tikzcircle[black, fill=white]{5pt},\overline{\tikzcircle[black, fill=black]{5pt}}\}\{\tikzcircle[black, fill=white]{5pt},\overline{\tikzcircle[black, fill=black]{5pt}}\}\{\tikzcircle[black, fill=white]{5pt},\overline{\tikzcircle[black, fill=black]{5pt}}\}\tikzcircle[black, fill=white]{5pt}\tikzcircle[black, fill=black]{5pt}\tikzcircle[black, fill=black]{5pt} \texttt{}{~~~~~~~~~~~~~~~~~~~~~~~~(6)}$$
$$\texttt{GHZ}(\tikzcircle[black, fill=black]{5pt}\tikzcircle[black, fill=black]{5pt}\tikzcircle[black, fill=black]{5pt}\tikzcircle[black, fill=black]{5pt}\tikzcircle[black, fill=white]{5pt}\tikzcircle[black, fill=black]{5pt}) = \{\tikzcircle[black, fill=white]{5pt},\overline{\tikzcircle[black, fill=black]{5pt}}\}\{\tikzcircle[black, fill=white]{5pt},\overline{\tikzcircle[black, fill=black]{5pt}}\}\{\tikzcircle[black, fill=white]{5pt},\overline{\tikzcircle[black, fill=black]{5pt}}\}\tikzcircle[black, fill=black]{5pt}\tikzcircle[black, fill=white]{5pt}\tikzcircle[black, fill=black]{5pt} \texttt{}{~~~~~~~~~~~~~~~~~~~~~~~~(7)}$$
$$\texttt{GHZ}(\tikzcircle[black, fill=black]{5pt}\tikzcircle[black, fill=black]{5pt}\tikzcircle[black, fill=black]{5pt}\tikzcircle[black, fill=black]{5pt}\tikzcircle[black, fill=black]{5pt}\tikzcircle[black, fill=white]{5pt}) = \{\tikzcircle[black, fill=white]{5pt},\overline{\tikzcircle[black, fill=black]{5pt}}\}\{\tikzcircle[black, fill=white]{5pt},\overline{\tikzcircle[black, fill=black]{5pt}}\}\{\tikzcircle[black, fill=white]{5pt},\overline{\tikzcircle[black, fill=black]{5pt}}\}\tikzcircle[black, fill=black]{5pt}\tikzcircle[black, fill=black]{5pt}\tikzcircle[black, fill=white]{5pt} \texttt{}{~~~~~~~~~~~~~~~~~~~~~~~~(8)}$$

The previous eight equations are equivalent to a misty state comprised 
of 64 combinations of six qubits each. We can group these by their last three
qubits as follows: $(1)$ with $(5)$, $(2)$ with $(6)$, $(3)$ with $(7)$, and $(4)$ with $(8)$. In each one of these pairs of groups the 
combinations that have an even number of $\tikzcircle[black, fill=black]{5pt}$'s cancel each other. We are then left with a combined 
misty state of 32 six-qubit combinations that have odd parity (an odd number of $\tikzcircle[black, fill=black]{5pt}$'s in them). This 
is half of the GHZ game. If we can show that the other half selects strings of even parity we're done.

\section{The Controlled-XZ Gate}

The first four gates in the $Y${\footnotesize -BOX} comprise what is known as the Controlled-XZ (\texttt{C-XZ}) gate: 

\begin{figure}[h]
    \centering
    \includegraphics[scale=0.19]{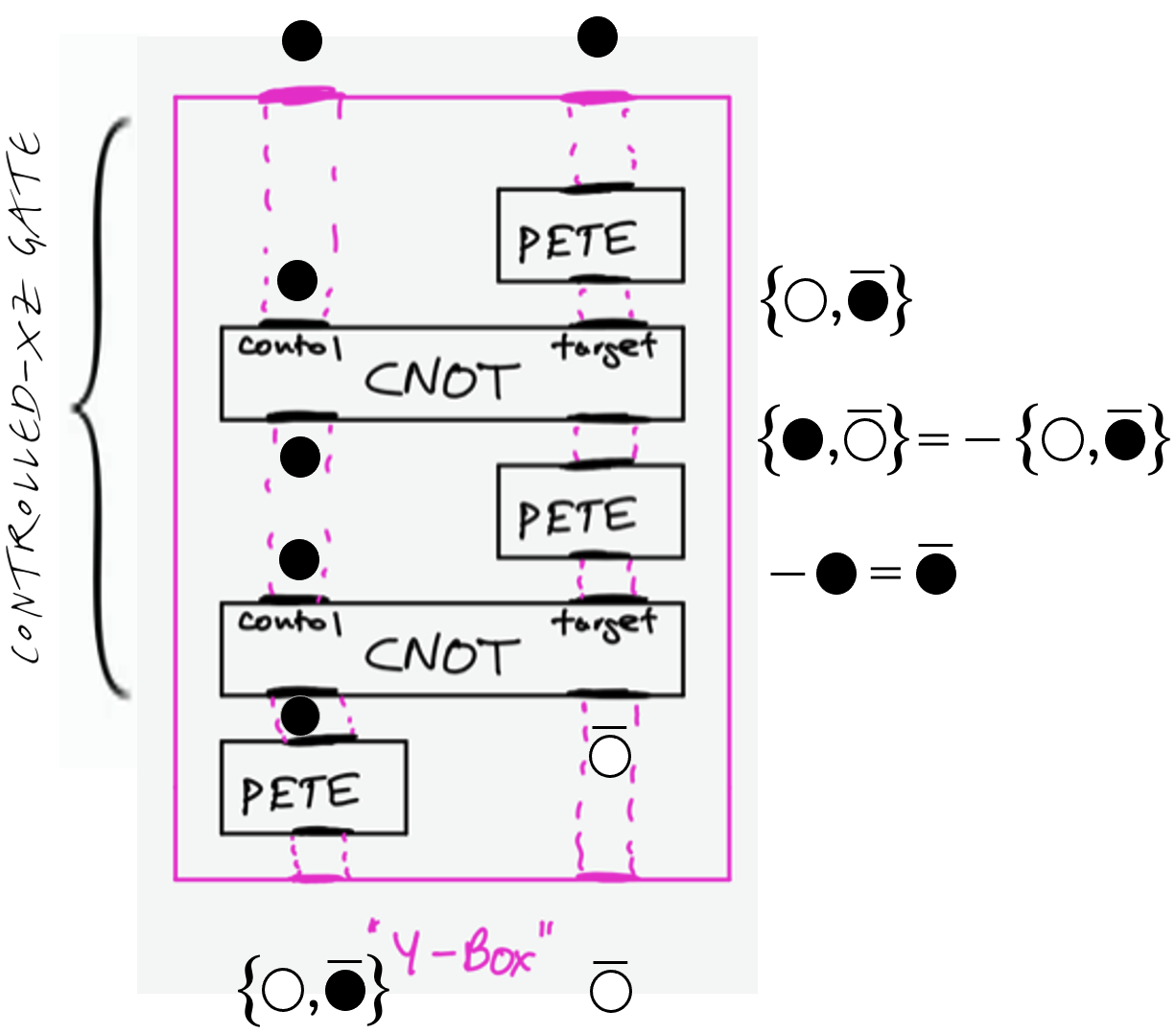}
    \label{fig:ghz-bw-05}
\end{figure}

With this we have (and we can see why the gate is named as such):

$$\texttt{C-XZ}(\tikzcircle[black, fill=white]{5pt}\tikzcircle[black, fill=white]{5pt}) = \tikzcircle[black, fill=white]{5pt}\tikzcircle[black, fill=white]{5pt}$$
$$\texttt{C-XZ}(\tikzcircle[black, fill=white]{5pt}\tikzcircle[black, fill=black]{5pt}) = \tikzcircle[black, fill=white]{5pt}\tikzcircle[black, fill=black]{5pt}$$
$$\texttt{C-XZ}(\tikzcircle[black, fill=black]{5pt}\tikzcircle[black, fill=white]{5pt}) = \tikzcircle[black, fill=black]{5pt}\tikzcircle[black, fill=black]{5pt}$$
$$\texttt{C-XZ}(\tikzcircle[black, fill=black]{5pt}\tikzcircle[black, fill=black]{5pt}) = \tikzcircle[black, fill=black]{5pt}\overline{\tikzcircle[black, fill=white]{5pt}}$$

As a reminder: instead of using complex numbers in a quantum  computer
we can always add a qubit to ``tag'' complex values. An explanation can be found in the paper by Lov K Grover and Terry Rudolph \cite{gr:01}. Now consider how the $S$ 
gate\footnote{$S = \begin{pmatrix} 1 & 0 \\ 0 & i \end{pmatrix}$}
gets implemented using the mapping from that paper and you will see it is implemented as a Controlled-XZ gate 
with target being the ``tagging'' (real/imaginary) qubit.
Measuring in the $X$ basis amounts to doing a Hadamard ({\footnotesize PETE} box) and then computational basis measurement. Measuring in the $Y$ basis can be done by first 
applying the $S$ followed by a Hadamard. 

\section{Calculations (Part II)}

From the previous section we have:
$$\texttt{Y-BOX}(\tikzcircle[black, fill=white]{5pt}\tikzcircle[black, fill=white]{5pt}) = \{\tikzcircle[black, fill=white]{5pt},\tikzcircle[black, fill=black]{5pt}\}\tikzcircle[black, fill=white]{5pt}$$
$$\texttt{Y-BOX}(\tikzcircle[black, fill=white]{5pt}\tikzcircle[black, fill=black]{5pt}) = \{\tikzcircle[black, fill=white]{5pt},\tikzcircle[black, fill=black]{5pt}\}\tikzcircle[black, fill=black]{5pt}$$
$$\texttt{Y-BOX}(\tikzcircle[black, fill=black]{5pt}\tikzcircle[black, fill=white]{5pt}) = \{\tikzcircle[black, fill=white]{5pt},\overline{\tikzcircle[black, fill=black]{5pt}}\}\tikzcircle[black, fill=black]{5pt}$$
$$\texttt{Y-BOX}(\tikzcircle[black, fill=black]{5pt}\tikzcircle[black, fill=black]{5pt}) = \{\tikzcircle[black, fill=white]{5pt},\overline{\tikzcircle[black, fill=black]{5pt}}\}\overline{\tikzcircle[black, fill=white]{5pt}}$$

Let's consider now the second part of the GHZ game with an input of $101$. In this case Alice and Carol must measure in the $Y$ basis and Bob in the $X$ basis. The
circuit is shown on the next page. We'll prove that all eight components in the misty state shared by the players result in strings with even parity. By symmetry we
don't need to check for the other inputs ($110$ and $011$) and the proof will have then been completed.

\newpage 

So we start calculating, again:

\begin{figure}[h]
    \centering
    \includegraphics[scale=0.17]{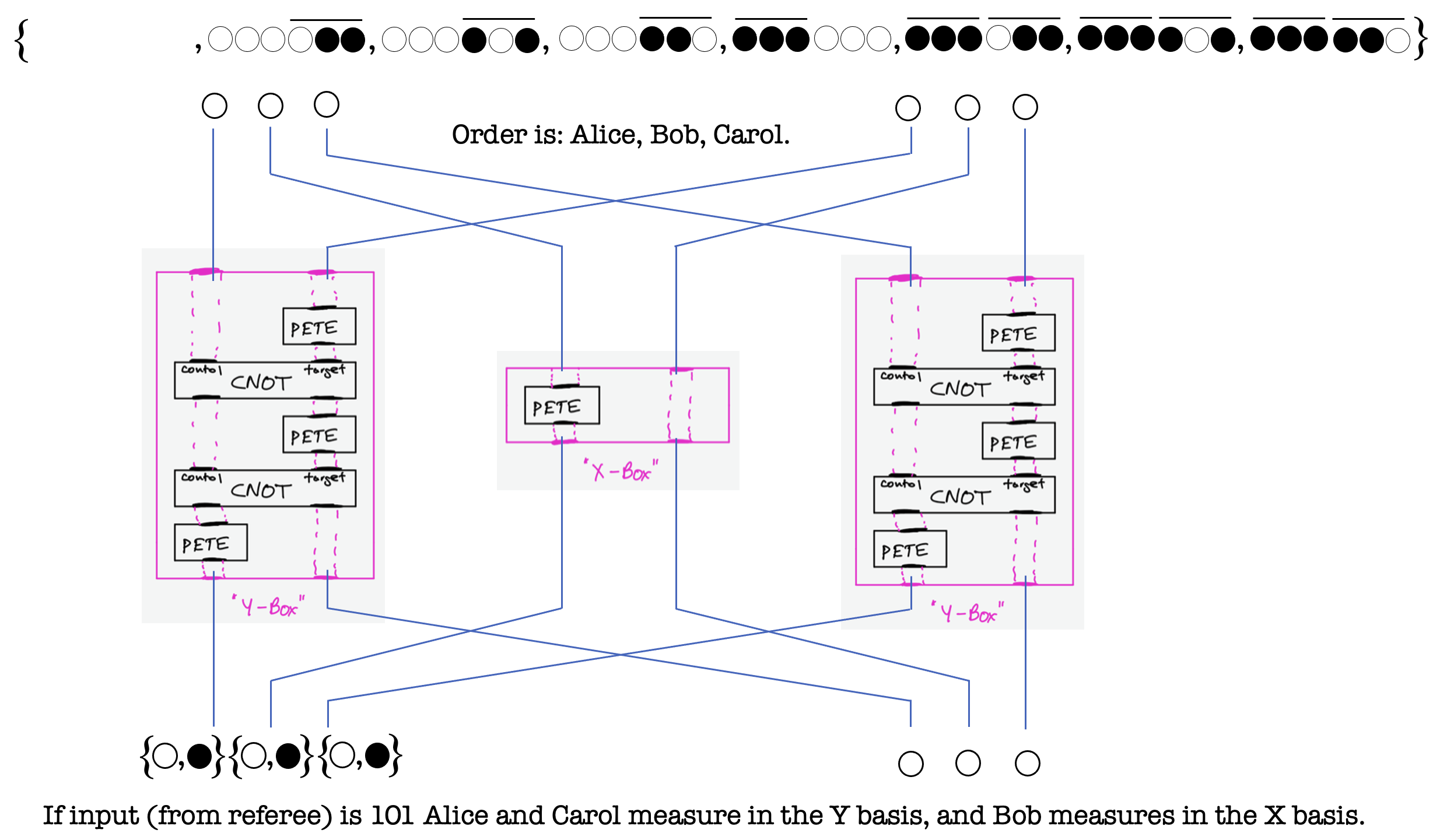}
    \label{fig:ghz-bw-06}
\end{figure}

And we have\footnote{For inputs $2$, $3$, $4$ and $5$ we again take the phase and place it on the fifth qubit (Bob's second). There is no phase
in the first input and phases cancel themselves out in the last three cases.}: 

$$\texttt{GHZ}(\tikzcircle[black, fill=white]{5pt}\tikzcircle[black, fill=white]{5pt}\tikzcircle[black, fill=white]{5pt}\tikzcircle[black, fill=white]{5pt}\tikzcircle[black, fill=white]{5pt}\tikzcircle[black, fill=white]{5pt}) = \{\tikzcircle[black, fill=white]{5pt},\tikzcircle[black, fill=black]{5pt}\}\{\tikzcircle[black, fill=white]{5pt},\tikzcircle[black, fill=black]{5pt}\}\{\tikzcircle[black, fill=white]{5pt},\tikzcircle[black, fill=black]{5pt}\}\tikzcircle[black, fill=white]{5pt}\tikzcircle[black, fill=white]{5pt}\tikzcircle[black, fill=white]{5pt} \texttt{}{~~~~~~~~~~~~~~~~~~~~~~~~(1)}$$
$$\texttt{GHZ}(\tikzcircle[black, fill=white]{5pt}\tikzcircle[black, fill=white]{5pt}\tikzcircle[black, fill=white]{5pt}\tikzcircle[black, fill=white]{5pt}\overline{\tikzcircle[black, fill=black]{5pt}}\tikzcircle[black, fill=black]{5pt}) = \{\tikzcircle[black, fill=white]{5pt},\tikzcircle[black, fill=black]{5pt}\}\{\tikzcircle[black, fill=white]{5pt},\tikzcircle[black, fill=black]{5pt}\}\{\tikzcircle[black, fill=white]{5pt},\tikzcircle[black, fill=black]{5pt}\}\tikzcircle[black, fill=white]{5pt}\overline{\tikzcircle[black, fill=black]{5pt}}\tikzcircle[black, fill=black]{5pt} \texttt{}{~~~~~~~~~~~~~~~~~~~~~~~~(2)}$$
$$\texttt{GHZ}(\tikzcircle[black, fill=white]{5pt}\tikzcircle[black, fill=white]{5pt}\tikzcircle[black, fill=white]{5pt}\tikzcircle[black, fill=black]{5pt}\overline{\tikzcircle[black, fill=white]{5pt}}\tikzcircle[black, fill=black]{5pt}) = \{\tikzcircle[black, fill=white]{5pt},\tikzcircle[black, fill=black]{5pt}\}\{\tikzcircle[black, fill=white]{5pt},\tikzcircle[black, fill=black]{5pt}\}\{\tikzcircle[black, fill=white]{5pt},\tikzcircle[black, fill=black]{5pt}\}\tikzcircle[black, fill=black]{5pt}\overline{\tikzcircle[black, fill=white]{5pt}}\tikzcircle[black, fill=black]{5pt} \texttt{}{~~~~~~~~~~~~~~~~~~~~~~~~(3)}$$
$$\texttt{GHZ}(\tikzcircle[black, fill=white]{5pt}\tikzcircle[black, fill=white]{5pt}\tikzcircle[black, fill=white]{5pt}\tikzcircle[black, fill=black]{5pt}\overline{\tikzcircle[black, fill=black]{5pt}}\tikzcircle[black, fill=white]{5pt}) = \{\tikzcircle[black, fill=white]{5pt},\tikzcircle[black, fill=black]{5pt}\}\{\tikzcircle[black, fill=white]{5pt},\tikzcircle[black, fill=black]{5pt}\}\{\tikzcircle[black, fill=white]{5pt},\tikzcircle[black, fill=black]{5pt}\}\tikzcircle[black, fill=black]{5pt}\overline{\tikzcircle[black, fill=black]{5pt}}\tikzcircle[black, fill=white]{5pt} \texttt{}{~~~~~~~~~~~~~~~~~~~~~~~~(4)}$$
$$\texttt{GHZ}(\tikzcircle[black, fill=black]{5pt}\tikzcircle[black, fill=black]{5pt}\tikzcircle[black, fill=black]{5pt}\tikzcircle[black, fill=white]{5pt}\overline{\tikzcircle[black, fill=white]{5pt}}\tikzcircle[black, fill=white]{5pt}) = \{\tikzcircle[black, fill=white]{5pt},\overline{\tikzcircle[black, fill=black]{5pt}}\}\{\tikzcircle[black, fill=white]{5pt},\overline{\tikzcircle[black, fill=black]{5pt}}\}\{\tikzcircle[black, fill=white]{5pt},\overline{\tikzcircle[black, fill=black]{5pt}}\}\tikzcircle[black, fill=black]{5pt}\overline{\tikzcircle[black, fill=white]{5pt}}\tikzcircle[black, fill=black]{5pt} \texttt{}{~~~~~~~~~~~~~~~~~~~~~~~~(5)}$$
$$\texttt{GHZ}(\tikzcircle[black, fill=black]{5pt}\tikzcircle[black, fill=black]{5pt}\tikzcircle[black, fill=black]{5pt}\tikzcircle[black, fill=white]{5pt}\tikzcircle[black, fill=black]{5pt}\tikzcircle[black, fill=black]{5pt}) = \{\tikzcircle[black, fill=white]{5pt},\overline{\tikzcircle[black, fill=black]{5pt}}\}\{\tikzcircle[black, fill=white]{5pt},\overline{\tikzcircle[black, fill=black]{5pt}}\}\{\tikzcircle[black, fill=white]{5pt},\overline{\tikzcircle[black, fill=black]{5pt}}\}\tikzcircle[black, fill=black]{5pt}\tikzcircle[black, fill=black]{5pt}\overline{\tikzcircle[black, fill=white]{5pt}}\texttt{}{~~~~~~~~~~~~~~~~~~~~~~~~(6)}$$
$$\texttt{GHZ}(\tikzcircle[black, fill=black]{5pt}\tikzcircle[black, fill=black]{5pt}\tikzcircle[black, fill=black]{5pt}\tikzcircle[black, fill=black]{5pt}\tikzcircle[black, fill=white]{5pt}\tikzcircle[black, fill=black]{5pt}) = \{\tikzcircle[black, fill=white]{5pt},\overline{\tikzcircle[black, fill=black]{5pt}}\}\{\tikzcircle[black, fill=white]{5pt},\overline{\tikzcircle[black, fill=black]{5pt}}\}\{\tikzcircle[black, fill=white]{5pt},\overline{\tikzcircle[black, fill=black]{5pt}}\}\overline{\tikzcircle[black, fill=white]{5pt}}\tikzcircle[black, fill=white]{5pt}\overline{\tikzcircle[black, fill=white]{5pt}} \texttt{}{~~~~~~~~~~~~~~~~~~~~~~~~(7)}$$
$$\texttt{GHZ}(\tikzcircle[black, fill=black]{5pt}\tikzcircle[black, fill=black]{5pt}\tikzcircle[black, fill=black]{5pt}\tikzcircle[black, fill=black]{5pt}\tikzcircle[black, fill=black]{5pt}\tikzcircle[black, fill=white]{5pt}) = \{\tikzcircle[black, fill=white]{5pt},\overline{\tikzcircle[black, fill=black]{5pt}}\}\{\tikzcircle[black, fill=white]{5pt},\overline{\tikzcircle[black, fill=black]{5pt}}\}\{\tikzcircle[black, fill=white]{5pt},\overline{\tikzcircle[black, fill=black]{5pt}}\}\overline{\tikzcircle[black, fill=white]{5pt}}\tikzcircle[black, fill=black]{5pt}\tikzcircle[black, fill=black]{5pt} \texttt{}{~~~~~~~~~~~~~~~~~~~~~~~~(8)}$$

The previous eight equations are equivalent to a misty state comprised 
of 64 combinations of six qubits each. We can group these by their last three
qubits as follows: $(1)$ with $(7)$, $(2)$ with $(8)$, $(3)$ with $(5)$, and $(4)$ with $(6)$. In each one of these pairs of groups the 
combinations that have an odd number of $\tikzcircle[black, fill=black]{5pt}$'s cancel each other. We are then left with a combined 
misty state of 32 six-qubit combinations that have even parity (an even number of $\tikzcircle[black, fill=black]{5pt}$'s in them). QED.

\section{Conclusion}

In \cite{semantics} the authors show how the GHZ game works using the misty states formalism from \cite{terry:01}. Their
straightforward approach there relies on four more colors in addition to the black and
white used in the book   \cite{terry:01} and they also need the singular fact that $i^2 = -1$. Here we show that it is possible for three people
to win  the GHZ game with just the standard real-valued misty state formalism, but each party will need to have two balls (qubits) instead of one. That
is, there will be six balls involved, rather than three. 

In fact the results of \cite{mmck:01} along with the proof that Toffoli plus Hadamard are universal 
means you can always play such a trick (i.e., expand the number of balls available, 
and you can implement some version of pretty much any quantum information problem). Of course it 
eventually becomes unwieldy to do so, and at some point students should be forced to bite the bullet
and learn\footnote{In \cite{semantics} it is
argued that relying on the misty states formalism in the early stages greatly simplifies this transition.}
full quantum theory!

In any event, to summarize the approach presented here, here's how the tree parties can win the game using the extra three balls. The first 
three balls are prepared in $\{\tikzcircle[black, fill=white]{5pt}\tikzcircle[black, fill=white]{5pt}\tikzcircle[black, fill=white]{5pt},-\tikzcircle[black, fill=black]{5pt}\tikzcircle[black, fill=black]{5pt}\tikzcircle[black, fill=black]{5pt}\}$ while the second three are prepared in 
$\{\tikzcircle[black, fill=white]{5pt}\tikzcircle[black, fill=white]{5pt}\tikzcircle[black, fill=white]{5pt},
-\tikzcircle[black, fill=white]{5pt}\tikzcircle[black, fill=black]{5pt}\tikzcircle[black, fill=black]{5pt},
-\tikzcircle[black, fill=black]{5pt}\tikzcircle[black, fill=white]{5pt}\tikzcircle[black, fill=black]{5pt},
-\tikzcircle[black, fill=black]{5pt}\tikzcircle[black, fill=black]{5pt}\tikzcircle[black, fill=white]{5pt}
\}$. Alice, Bob and Carol take one ball from each set (the states are permutation invariant). In the GHZ game
either all three are asked the ``$X$'' question, or one is asked ``$X$'' and the other two are asked ``$Y$''. Each party 
locally has two setups that we have labeled $X${\footnotesize -BOX} and $Y${\footnotesize -BOX}; their structure and operation have been presented earlier in detail. 
Depending on what they are asked they drop their
two balls through one setup or the other. The ball they took from the first set goes through the $X${\footnotesize -BOX}, the ball
they took from the second set goes through their $Y${\footnotesize -BOX}. Then they observe the color of their first (left) ball and report it
as $\pm 1$ as normal. 

Why does it work? If all three are asked ``$X$'' then they are all just doing the standard Hadamard ({\footnotesize PETE}) onto the misty state
for the first three bits and this produces a superposition of terms that only have odd parity. What is going on in the $Y${\footnotesize -BOX} is
a bit trickier. The first four boxes implement a Controlled-XZ (\texttt{C-XZ}) gate, with the first ball as the control and the second one as the target. 
Now in \cite{mmck:01} the authors point out that applying (\texttt{XZ}) to any 
two qubits of the state
$\{\tikzcircle[black, fill=white]{5pt}\tikzcircle[black, fill=white]{5pt}\tikzcircle[black, fill=white]{5pt},
\overline{\tikzcircle[black, fill=white]{5pt}\tikzcircle[black, fill=black]{5pt}\tikzcircle[black, fill=black]{5pt}},
\overline{\tikzcircle[black, fill=black]{5pt}\tikzcircle[black, fill=white]{5pt}\tikzcircle[black, fill=black]{5pt}},
\overline{\tikzcircle[black, fill=black]{5pt}\tikzcircle[black, fill=black]{5pt}\tikzcircle[black, fill=white]{5pt}}
\}$ converts this state to the negative of itself\footnote{See details of how this occurs in the attached Appendix.}. 
With a bit of thought the reader will be able to convince themselves that because we 
started with $\{\tikzcircle[black, fill=white]{5pt}\tikzcircle[black, fill=white]{5pt}\tikzcircle[black, fill=white]{5pt},\overline{\tikzcircle[black, fill=black]{5pt}\tikzcircle[black, fill=black]{5pt}\tikzcircle[black, fill=black]{5pt}}\}$ on the control balls, once two of the parties apply this controlled-(XZ) the state of 
the first three balls will get converted to $\{\tikzcircle[black, fill=white]{5pt}\tikzcircle[black, fill=white]{5pt}\tikzcircle[black, fill=white]{5pt},\tikzcircle[black, fill=black]{5pt}\tikzcircle[black, fill=black]{5pt}\tikzcircle[black, fill=black]{5pt}\}$ (the first three balls return to a product state with the second three by the end
of the $Y${\footnotesize -BOX}'es) and when the final Hadamards are now applied this state becomes a superposition of even parity terms.

\section*{Appendix}

The key to GHZ is that if you begin with the misty state $$\{ \tikzcircle[black, fill=white]{5pt}\tikzcircle[black, fill=white]{5pt}\tikzcircle[black, fill=white]{5pt}, \overline{\tikzcircle[black, fill=black]{5pt}\tikzcircle[black, fill=black]{5pt}\tikzcircle[black, fill=black]{5pt}}\}$$ and send all three balls through 
{\footnotesize PETE} boxes then 
the output mist is $$\{ \tikzcircle[black, fill=white]{5pt}\tikzcircle[black, fill=white]{5pt}\tikzcircle[black, fill=black]{5pt}, \tikzcircle[black, fill=white]{5pt}\tikzcircle[black, fill=black]{5pt}\tikzcircle[black, fill=white]{5pt}, \tikzcircle[black, fill=black]{5pt}\tikzcircle[black, fill=white]{5pt}\tikzcircle[black, fill=white]{5pt}, \tikzcircle[black, fill=black]{5pt}\tikzcircle[black, fill=black]{5pt}\tikzcircle[black, fill=black]{5pt}\}$$ while if you begin with the misty state 
$$\{ \tikzcircle[black, fill=white]{5pt}\tikzcircle[black, fill=white]{5pt}\tikzcircle[black, fill=white]{5pt}, \tikzcircle[black, fill=black]{5pt}\tikzcircle[black, fill=black]{5pt}\tikzcircle[black, fill=black]{5pt}\}$$ and send all three balls through 
{\footnotesize PETE}
boxes then the output is 
$$\{ \tikzcircle[black, fill=white]{5pt}\tikzcircle[black, fill=white]{5pt}\tikzcircle[black, fill=white]{5pt}, \tikzcircle[black, fill=white]{5pt}\tikzcircle[black, fill=black]{5pt}\tikzcircle[black, fill=black]{5pt}, \tikzcircle[black, fill=black]{5pt}\tikzcircle[black, fill=white]{5pt}\tikzcircle[black, fill=black]{5pt}, \tikzcircle[black, fill=black]{5pt}\tikzcircle[black, fill=black]{5pt}\tikzcircle[black, fill=white]{5pt}\}$$ In the first case the parity of the outputs is odd, in the second is even. Therefore, assuming the parties being sharing $$\{ \tikzcircle[black, fill=white]{5pt}\tikzcircle[black, fill=white]{5pt}\tikzcircle[black, fill=white]{5pt}, \overline{\tikzcircle[black, fill=black]{5pt}\tikzcircle[black, fill=black]{5pt}\tikzcircle[black, fill=black]{5pt}}\}$$ what we want is a way for them to locally convert this to $$\{ \tikzcircle[black, fill=white]{5pt}\tikzcircle[black, fill=white]{5pt}\tikzcircle[black, fill=white]{5pt}, \tikzcircle[black, fill=black]{5pt}\tikzcircle[black, fill=black]{5pt}\tikzcircle[black, fill=black]{5pt}\}$$ in the event that they get asked ``one $X$, two $Y$'s''. In regular quantum mechanics this can be achieved by each party that gets asked a 
$Y$ question
first doing the $S$ gate on their qubit before they do the Hadamard. That then leaves 
$\tikzcircle[black, fill=white]{5pt}\tikzcircle[black, fill=white]{5pt}\tikzcircle[black, fill=white]{5pt}$ invariant but it converts 
$\overline{ \tikzcircle[black, fill=black]{5pt}\tikzcircle[black, fill=black]{5pt}\tikzcircle[black, fill=black]{5pt} }$ to 
$\tikzcircle[black, fill=black]{5pt}\tikzcircle[black, fill=black]{5pt}\tikzcircle[black, fill=black]{5pt} $. But in the regular misty state formalism from 
\cite{terry:01} the $S$ gates aren't built in.

The way we are going to flip $\overline{ \tikzcircle[black, fill=black]{5pt}\tikzcircle[black, fill=black]{5pt}\tikzcircle[black, fill=black]{5pt} }$ (but
leave $ \tikzcircle[black, fill=white]{5pt}\tikzcircle[black, fill=white]{5pt}\tikzcircle[black, fill=white]{5pt} $ as it is) is to bring a completely new
three ball mist, with one ball to each player. For the moment lets just denote this mist $\{ \texttt{M} \}$ for ``magic''. In the misty formalism the initial
state is then $$\{ \tikzcircle[black, fill=white]{5pt}\tikzcircle[black, fill=white]{5pt}\tikzcircle[black, fill=white]{5pt}, \overline{\tikzcircle[black, fill=black]{5pt}\tikzcircle[black, fill=black]{5pt}\tikzcircle[black, fill=black]{5pt}}\}\{\texttt{M}\}$$ which is the same as (note the brackets carefully) $$\{ \tikzcircle[black, fill=white]{5pt}\tikzcircle[black, fill=white]{5pt}\tikzcircle[black, fill=white]{5pt}\{\texttt{M}\}, \overline{\tikzcircle[black, fill=black]{5pt}\tikzcircle[black, fill=black]{5pt}\tikzcircle[black, fill=black]{5pt}}\{\texttt{M}\}\}$$ Let's think about what we want to happen when two of the parties drop their balls through a
$Y${\footnotesize -BOX}: we want $\tikzcircle[black, fill=white]{5pt}\tikzcircle[black, fill=white]{5pt}\tikzcircle[black, fill=white]{5pt}\{\texttt{M}\}$ to remain the same but
the $\overline{\tikzcircle[black, fill=black]{5pt}\tikzcircle[black, fill=black]{5pt}\tikzcircle[black, fill=black]{5pt}}\{\texttt{M}\}$ to evolve to  
$\tikzcircle[black, fill=black]{5pt}\tikzcircle[black, fill=black]{5pt}\tikzcircle[black, fill=black]{5pt}\{\texttt{M}\}$. 

Now when we are using the first three balls as the controls in the 
$Y${\footnotesize -BOX}-es, and the other three $\{\texttt{M}\}$-balls as the targets, 
the $\tikzcircle[black, fill=white]{5pt}\tikzcircle[black, fill=white]{5pt}\tikzcircle[black, fill=white]{5pt}$ 
part of the mist for the first three balls just leaves $\{\texttt{M}\}$ as $\{\texttt{M}\}$. But the
$\tikzcircle[black, fill=black]{5pt}\tikzcircle[black, fill=black]{5pt}\tikzcircle[black, fill=black]{5pt} $
part of the mist for the first three balls will mean that two of the balls of the 
$\{\texttt{M}\}$
mist will see an (\texttt{XZ}) box, which (by definition, or by construction) should take $\{\texttt{M}\}$ to  $\overline{\{\texttt{M}\}}$. And then we have: 
$$\{ \tikzcircle[black, fill=white]{5pt}\tikzcircle[black, fill=white]{5pt}\tikzcircle[black, fill=white]{5pt}~\{\texttt{M}\}, \overline{\tikzcircle[black, fill=black]{5pt}\tikzcircle[black, fill=black]{5pt}\tikzcircle[black, fill=black]{5pt}}~\overline{\{\texttt{M}\}}\} = \{ \tikzcircle[black, fill=white]{5pt}\tikzcircle[black, fill=white]{5pt}\tikzcircle[black, fill=white]{5pt}\{\texttt{M}\}, \tikzcircle[black, fill=black]{5pt}\tikzcircle[black, fill=black]{5pt}\tikzcircle[black, fill=black]{5pt}\{\texttt{M}\}\} = \{ \tikzcircle[black, fill=white]{5pt}\tikzcircle[black, fill=white]{5pt}\tikzcircle[black, fill=white]{5pt}, \tikzcircle[black, fill=black]{5pt}\tikzcircle[black, fill=black]{5pt}\tikzcircle[black, fill=black]{5pt}\}\{\texttt{M}\}$$
And now the first three balls are in the correct state for winning the game. The upshot is that the magic all comes about because it is possible
to find a suitable $\{\texttt{M}\}$, namely $$\{\texttt{M}\} = \{ \tikzcircle[black, fill=white]{5pt}\tikzcircle[black, fill=white]{5pt}\tikzcircle[black, fill=white]{5pt}, \overline{\tikzcircle[black, fill=white]{5pt}\tikzcircle[black, fill=black]{5pt}\tikzcircle[black, fill=black]{5pt}}, \overline{\tikzcircle[black, fill=black]{5pt}\tikzcircle[black, fill=white]{5pt}\tikzcircle[black, fill=black]{5pt}}, \overline{\tikzcircle[black, fill=black]{5pt}\tikzcircle[black, fill=black]{5pt}\tikzcircle[black, fill=white]{5pt}}\}$$ which the reader can verify (perhaps somewhat painfully) by explicit calculation. 
We show that below.

Let us take $$\{\texttt{M}\} = \{ \tikzcircle[black, fill=white]{5pt}\tikzcircle[black, fill=white]{5pt}\tikzcircle[black, fill=white]{5pt}, \overline{\tikzcircle[black, fill=white]{5pt}\tikzcircle[black, fill=black]{5pt}\tikzcircle[black, fill=black]{5pt}}, \overline{\tikzcircle[black, fill=black]{5pt}\tikzcircle[black, fill=white]{5pt}\tikzcircle[black, fill=black]{5pt}}, \overline{\tikzcircle[black, fill=black]{5pt}\tikzcircle[black, fill=black]{5pt}\tikzcircle[black, fill=white]{5pt}}\}$$ The state is permutation invariant. What we mean by $-\{\texttt{M}\} = \overline{\{\texttt{M}\}} = \{\overline{\texttt{M}}\} = \{\texttt{-M}\}$ is: $$\{ \overline{\tikzcircle[black, fill=white]{5pt}\tikzcircle[black, fill=white]{5pt}\tikzcircle[black, fill=white]{5pt}}, \tikzcircle[black, fill=white]{5pt}\tikzcircle[black, fill=black]{5pt}\tikzcircle[black, fill=black]{5pt}, \tikzcircle[black, fill=black]{5pt}\tikzcircle[black, fill=white]{5pt}\tikzcircle[black, fill=black]{5pt}, \tikzcircle[black, fill=black]{5pt}\tikzcircle[black, fill=black]{5pt}\tikzcircle[black, fill=white]{5pt}\}$$ This is still an even parity state, but with just different negative signs. 

Consider the $Y${\footnotesize -BOX}. If you look at what happens to the second ball if the first ball is white, then you see that it will go through just two {\footnotesize PETE} boxes, which takes 
$\tikzcircle[black, fill=white]{5pt}$ into $\tikzcircle[black, fill=white]{5pt}$, and $\tikzcircle[black, fill=black]{5pt}$ into $\tikzcircle[black, fill=black]{5pt}$, i.e. the same as doing nothing.
But if the first ball is black then the C-NOT boxes mean the second ball will experience {\footnotesize PETE} then NOT then {\footnotesize PETE} then NOT. You can check that the combination of those 
4 boxes does the transformation $\tikzcircle[black, fill=white]{5pt} \rightarrow \tikzcircle[black, fill=black]{5pt}$ and 
$\tikzcircle[black, fill=black]{5pt} \rightarrow \overline{\tikzcircle[black, fill=white]{5pt}}$, respectively. We could call
this a single (\texttt{XZ}) box. 

Consider what happens to $\{\texttt{M}\}$ if you send two of the balls (say the first two) each through an (\texttt{XZ}) box:
$$\{\texttt{M}\} = \{ \tikzcircle[black, fill=white]{5pt}\tikzcircle[black, fill=white]{5pt}\tikzcircle[black, fill=white]{5pt}, \overline{\tikzcircle[black, fill=white]{5pt}\tikzcircle[black, fill=black]{5pt}\tikzcircle[black, fill=black]{5pt}}, \overline{\tikzcircle[black, fill=black]{5pt}\tikzcircle[black, fill=white]{5pt}\tikzcircle[black, fill=black]{5pt}}, \overline{\tikzcircle[black, fill=black]{5pt}\tikzcircle[black, fill=black]{5pt}\tikzcircle[black, fill=white]{5pt}}\}$$ will evolve to:
$$\{ \tikzcircle[black, fill=black]{5pt}\tikzcircle[black, fill=black]{5pt}\tikzcircle[black, fill=white]{5pt}, \overline{\tikzcircle[black, fill=black]{5pt}\overline{\tikzcircle[black, fill=white]{5pt}}\tikzcircle[black, fill=black]{5pt}}, \overline{\overline{\tikzcircle[black, fill=white]{5pt}}\tikzcircle[black, fill=black]{5pt}\tikzcircle[black, fill=black]{5pt}}, \overline{\overline{\tikzcircle[black, fill=white]{5pt}}\overline{\tikzcircle[black, fill=white]{5pt}}\tikzcircle[black, fill=white]{5pt}}\}$$ which is the same as:
$$\{ \tikzcircle[black, fill=black]{5pt}\tikzcircle[black, fill=black]{5pt}\tikzcircle[black, fill=white]{5pt}, \tikzcircle[black, fill=black]{5pt}\tikzcircle[black, fill=white]{5pt}\tikzcircle[black, fill=black]{5pt}, \tikzcircle[black, fill=white]{5pt}\tikzcircle[black, fill=black]{5pt}\tikzcircle[black, fill=black]{5pt}, \overline{\tikzcircle[black, fill=white]{5pt}\tikzcircle[black, fill=white]{5pt}\tikzcircle[black, fill=white]{5pt}}\}$$ which is: $$\{\overline{\tikzcircle[black, fill=white]{5pt}\tikzcircle[black, fill=white]{5pt}\tikzcircle[black, fill=white]{5pt}}  ,  \tikzcircle[black, fill=white]{5pt}\tikzcircle[black, fill=black]{5pt}\tikzcircle[black, fill=black]{5pt}, 
\tikzcircle[black, fill=black]{5pt}\tikzcircle[black, fill=white]{5pt}\tikzcircle[black, fill=black]{5pt},
\tikzcircle[black, fill=black]{5pt}\tikzcircle[black, fill=black]{5pt}\tikzcircle[black, fill=white]{5pt}\} = \overline{  \{ \tikzcircle[black, fill=white]{5pt}\tikzcircle[black, fill=white]{5pt}\tikzcircle[black, fill=white]{5pt}, \overline{\tikzcircle[black, fill=white]{5pt}\tikzcircle[black, fill=black]{5pt}\tikzcircle[black, fill=black]{5pt}}, \overline{\tikzcircle[black, fill=black]{5pt}\tikzcircle[black, fill=white]{5pt}\tikzcircle[black, fill=black]{5pt}}, \overline{\tikzcircle[black, fill=black]{5pt}\tikzcircle[black, fill=black]{5pt}\tikzcircle[black, fill=white]{5pt}}\} } = \overline{\{ \texttt{M} \}}$$ 
So, one more time,  when we are using the first three balls as the controls in the $Y${\footnotesize -BOX}-es, and the other three $\{\texttt{M}\}$-balls as the targets, the 
$\tikzcircle[black, fill=white]{5pt}\tikzcircle[black, fill=white]{5pt}\tikzcircle[black, fill=white]{5pt}$ part of the mist for the first three balls just leaves $\{\texttt{M}\}$ as 
$\{\texttt{M}\}$. But the $\tikzcircle[black, fill=black]{5pt}\tikzcircle[black, fill=black]{5pt}\tikzcircle[black, fill=black]{5pt}$ part
of the mist for the first three balls will mean that two of the balls of the $\{\texttt{M}\}$ mist will see an  (\texttt{XZ}) box, and as we just saw that takes 
$\{\texttt{M}\}$ to $\overline{\{\texttt{M}\}}$.


\begin{thebibliography}{9}

\bibitem{terry:01} Terry Rudolph (2017) \emph{Q is for Quantum}. Published by Terence Rudolph (July 19, 2017).


\bibitem{ghz:01} Jian Wei-Pan, Dik Bouwmeester, Matthew Daniell, Harald Weinfurter and Anton Zeilinger, ``Experimental 
test of quantum nonlocality in three-photon Greenberger-Horne-Zeilinger entanglement.'' Letters to Nature (Vol 403, Feb. 2000) Available at \url{http://web.stanford.edu/~oas/SI/QM/notes/GHZreference2.pdf}. Accessed: July 8, 2026.

\bibitem{mmck:01} Matthew McKague, Michele Mosca and Nicolas Gisin, ``Simulating quantum systems using real Hilbert spaces'' (October 2008). Available at  \url{https://arxiv.org/pdf/0810.1923.pdf}. Accessed: July 8, 2026.

\bibitem{gr:01} Lov K Grover and Terry Rudolph, ``A 2 rebit gate universal for quantum computing'' (October 2002) Available
at: \url{https://arxiv.org/abs/quant-ph/0210187}. Accessed: July 8, 2026.

\bibitem{semantics} 
Dan-Adrian German; ``A Term-Rewriting Semantics for Pure Quantum States'' (5 July, 2026). 
Available at: \url{https://arxiv.org/abs/2607.06584}. Accessed: July 8, 2026.

\bibitem{terry:02} Terry Rudolph (2017) Website, videos and other useful auxiliary materials for the book \emph{Q is for Quantum}. Available at: \url{https://www.qisforquantum.org/}. Accessed: July 8, 2026.
\end{thebibliography}
\end{document}